\documentclass[square]{ws-procs11x85}
\usepackage{balance} %necessary when used with \twocolumn

\newcommand{\pc}{{\rm pc}}

\newcommand{\mas}{{\rm mas}}
\newcommand{\degr}{\mbox{$^\circ$}}
\newcommand{\itprod}{\! \cdot \!}        % inner tensor (dot) product
\newcommand{\ilos}{_{\rm los}}
\newcommand{\los}{\hat n\ilos}     % line-of-sight direction
\newcommand{\cross}{\times}              % cross product
\begin{document}

\twocolumn[
\title{MHD models and synthetic synchrotron maps for the jet of M87}

\author{J. Gracia}

\address{Dublin Institute for Advanced Studies, 31 Fitzwilliam Place, Dublin 2,
  Ireland\\E-mail: jgracia@cp.dias.ie}

\author{S. Bogovalov, K. Tsinganos}

%%%%%%%%%%%%%%%%%%%%%%%%%%%%%%%%%%%%%%%%%%%%%%%%%%%%%%%%%%%%%%%%%%%%%%%%%
% You may repeat \author \address as often as necessary                 %
%%%%%%%%%%%%%%%%%%%%%%%%%%%%%%%%%%%%%%%%%%%%%%%%%%%%%%%%%%%%%%%%%%%%%%%%%

\begin{abstract}
We present a self-consistent MHD model for the jet
of M87. The model consist of two distinct zones: an inner relativistic
outflow, which we identify with the observed jet, and an outer cold
disk-wind. While the former does not self-collimate efficiently due to
its high effective inertia, the latter fulfills all the conditions for
efficient collimation by the magneto-centrifugal mechanism. Given the
right balance between the effective inertia of the inner flow and the
collimation efficiency of the outer disk wind, the relativistic flow
is magnetically confined into a well collimated beam for a wide range
of parameters and matches the measurements of the opening angle of
M87 over several orders of magnitude in spatial extent. 

In the second part of this work, we present synthetic synchrotron
emission maps for our MHD models. In principle the two-zone model can
reproduce the morphological structure seen in radio observations, as
central-peaked profiles across the jet close the the source,
limb-bright further down the jet, and a bright knot close to the
position of HST-1. However it is difficult to reconcile all features
into a single set of parameters.

\end{abstract}
%\keywords{Keyword1; Keyword2}
%\vskip12pt  % insert '\vskip12pt' while using '\twocolumn' command
\vskip28pt % if there is no keywords
]

\bodymatter

\section{Introduction}
Since its discovery by Curtis the jet of M87 is the classical
prototype for extragalactic jets.  Therefore, it is an ideal candidate
for testing specific jet formation mechanisms.  The jet and its hot
spots have been systematically studied across the electromagnetic
spectrum from the radio to X-rays, both, with ground-based
observations and from satellites.
% (for a review see e.g. \cite{Biretta96}). 
The jet is clearly detected at mm wavelength with
a resolution of 0.009 \pc{} out to distances of about 2
\mas{} from the core. The initial opening angle is
approximately 60\degr{} on scales of about 0.04 \pc{} and
decreases rapidly until reaching 10\degr{} at a distance of 4 \pc{}
from the core \cite{Biretta+02}. These observations suggest, that the
jet of M87 is rather slowly collimated across a length of several
parsec. 
%This scale is significantly larger than
%the radius of the black hole or any of the characteristic orbits,
%e.g. the last-stable orbit, but well below the size of the
%accretion disk, which can be as large as 20 \pc{}.

The prevailing paradigm for jet formation and collimation is magnetic
self-collimation by the Blandford \& Payne process
\cite{BP82}. 
%This view is supported by observational evidence
%\cite{Gabuzda+04} for a non-vanishing toroidal magnetic field
%component.
However, relativistic effects have been shown to decrease
the collimation efficiency \cite{BT99, TB00}.
%, i.e. for a given magnetic field
%configuration at the base of the jet (or rotator efficiency), the
%fraction of total mass- and magnetic flux that is asymptotically
%cylindrically collimated is lower for a relativistic flow than for a
%non-relativistic flow \cite{BT99, TB00}, due to its large effective
%momentum which counter-acts the pinching by the toroidal magnetic
%field. 
Not only is the mass flux fraction very low, but the final
opening angle is larger for relativistic outflows due to the
decollimating effect of the electric field and the effective inertia
of the plasma \cite{B01}, which both counter-act the pinching by the
toroidal magnetic field.

In a series of papers \cite{GTB05, TB05, TB02} tackled this problem
and suggested a two-zone model. The model consists of an inner
relativistic outflow, which is identified with the observed jet, and
an outer non-relativistic disk-wind. 
%While the inner relativistic jet
%is not expected to collimate well through magnetic self-collimation,
%the very same process operates very efficiently in the outer
%non-relativistic disk-wind. It is expected, that at least for a part
%of the available parameter space, collimation in the outer disk-wind
%is so efficient, that is might resist the decollimating inertia of the
%inner relativistic plasma and channel the jet into a narrowly confined
%beam. 
\cite{GTB05} could show, that such two-zone models could easily
account for the narrow appearance of the beam of extragalactic jets by
reproducing the observational measurement of the opening angle
distribution as a function of angular distance from the core as collected by
\cite{Biretta+02}. 
%Since in these models the jet -- identified with
%the relativistic inner outflow -- is strictly speaking not
%magnetically self-collimated, but rather confined by the outer
%disk-wind, the authors prefer to talk of collimation by magnetic
%confinement.

%However, \cite{GTB05} did not fit the opening angle distribution with
%a unique model. Instead o whole range of parameters reproduces the
%observations with similar accuracy. This may indicate, that the model
%is too complicated and could actually be simplified by finding, so far
%hidden relations, between the parameters. Indeed, as already
%indicated, the magnetic confiment efficiency is a function of the
%relative efficiencies of the decollimation of the inner relativistic
%flow and the collimation of the outer disk-wind, ie a combination of
%effective inertia and pinching force.

In deriving the opening angle of their model, \cite{GTB05} made a
simplifying but crucial assumption. They identified the observed jet
with the inner relativistic outflow of their two-zone model. More
specifically, the boundary of the jet was assumed to coincide with the
shape of a specific fieldline $\Psi_\alpha$
%, the one fieldline, which
separating the relativistic inner region from the non-relativistic
outer disk-wind at the base, or the launching surface, of the
outflow. So, the observed opening angle was strictly speaking fitted
by the shape of a single magnetic fieldline.

%How well justified is this assumption? The separating fieldline
%$\Psi_\alpha$ perfectly separates the relativistic from the
%non-relativistic outflow. Inside of $\Psi_\alpha$ the plasma is highly
%relativistic, both in terms of its bulk Gamma factor, $\Gamma >> 1$,
%and of its thermal energy, $T \le mc^2$, while outside the plasma is
%cold, $T<<mc^2$, and moving at non-relativistic speeds, $\Gamma =
%1$. Also, the magnetic field strength peaks close to the separating
%fieldline, where fieldlines (and poloidal flowlines) of the inner
%outflow are strongly compressed laterally and confined to a narrow
%sheet by the fieldlines of the outer disk-wind thus forming a natural
%interface between both zones.

%However, observations do not meassure the plasma state directly, i.e.
%in terms of velocity, temperature or magnetic field strength. Instead,
%they register photon flux as a function of position on the plane of
%the sky. As such, 
From an observational point of view, the width of the
jet is defined by the emission dropping below the detection limit or a
small fraction of the luminosity of the ridge line of the jet. So the
question is still, how well did \cite{GTB05} measure the width of
the jet in terms of observational quantities?

The purpose of this paper is to answer exactly this question by
adopting the point of view of an observer. Assuming, that the main
radiation mechanism is synchrotron emission, we translate the MHD
model into a synthetic emission map and measure the width of the jet
using only this data. 
%The gains are twofold. Firstly, we might indeed
%establish, that two-zone models can reproduce the {\em observed} width
%of the jet as a function of distance from the core. Secondly, since
%the data basis is much larger -- 2-dimensional emission and
%polarization maps versus 1-dimensional opening angle distribution --
%be might actually be able to fully invert the problem and determine a
%{\em unique set} of physical parameters for the jet of M87 within the
%framework of two-zone, i.e. collimation by magnetic confinement,
%models.

\section{MHD model}

\begin{figure}[t]
\center
\centerline{\psfig{figure=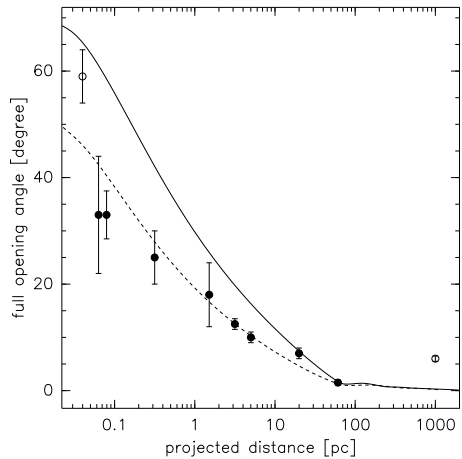,width=6truecm}}
\caption{Comparison of the opening angle calculated from model 1 ({\em
solid lines}) and model 2 ({\em dashed line}), respectively, and the
observational data for the jet of M87 ({\em data points}).}
\label{fig:opening_angle}
\end{figure}

\begin{figure}[t]
\center
\centerline{\psfig{figure=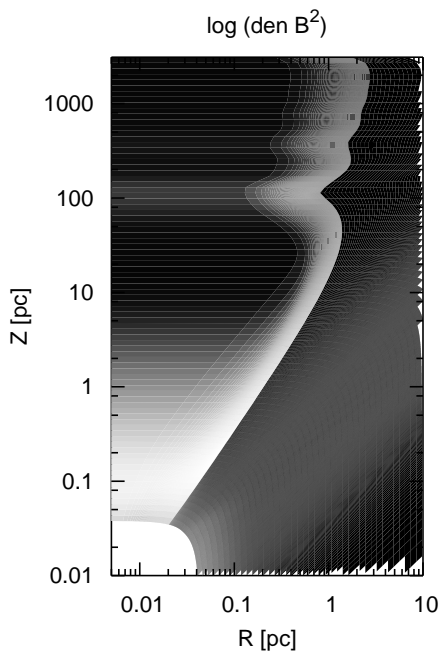,width=6truecm}}
\caption{Estimate of synchrotron emissivity through $\rho \, B^2$ for model 1.}
\label{fig:synch_estimate}
\end{figure}

We adopt the model and notation of \cite{GTB05} and refer the reader
to that paper for details. There you will also find a schematic of the
model in Fig. 4.

The model consist of two distinct zones, an inner outflow,
which is dominated by relativistic dynamics, and an outer
non-relativistic outflow. Both outflows originate from a spherical
launching surface
%  located at a distance $r_0$ from the black hole. The
%launching surface is 
threaded by a radial magnetic field. The two zones 
are separated by a specific fieldline $\Psi_\alpha$, where
$2\alpha$ is the initial angular width, or opening angle, of the inner
relativistic outflow.
%In the following, we will refer to these two
%distinct zones by relativistic jet and (non-relativistic) disk-wind,
%respectively.
%
%This two-zone model is motivated by a similar two-component structure
%of the underlying accretion disk consisting of an outer standard disk
%\citep{SS73} and an inner hot plasma, which could be either an
%advection dominated accretion flow \citep{NY94, PA97, GPKC03} or the
%final plunging region near the black hole, where relativistic dynamics
%dominates through e.g. frame-dragging or the Blandford-Znajek process.
%
We impose two different set of boundary conditions in the two distinct
zone along the launching surface. 
%If the launching surface is located
%beyond the fast magnetosonic surface, i.e. in the hyperbolic MHD
%regime, the steady-state problem reduces to an initial value
%Cauchy-type problem and can be integrated directly in terms conserved
%integrals of motion as described in \citet{TB02}. We stress, that
Beyond the launching surface we solve the problem self-consistently,
including the magnetic field structure, as a function of the boundary
values alone. 
%However, we do not solve the problem inside the
%launching surface, which is a much more complicated exercise. A
%self-consistent solution of the full problem needs to take into
%account the dynamics of the accretion flow and is beyond the scope of
%this paper.

This procedure yields a set of quantities as a function of space in
the comoving frame of the jet. GBT05 used the shape of the separating
fieldline $\Psi_\alpha$ to fit the observed opening angle as a
function of distance from the core as collected by \cite{Biretta+02}.
Unfortunately, the parameter space of the MHD model is degenerated in
the sense, that very different set of parameters yield equally good
fits or even almost identical opening angle distributions. Only 3-4 of
the formally 7 parameters of the model seem to be independent. 

Typically, models that fit the observed opening angle well are
moderately relativistic, both in terms of the initial outflow velocity
$\Gamma \sim 2$ and the initial plasma internal energy $T \sim 3
mc^2$.  The plasma velocity increases along the flow to values up to
$\Gamma \sim 6-8$. However, the plasma may decelerate and reaccelerate
sharply if recollimation takes place. Strong gradients of the plasma
velocity may generally be present across the jet. It is therefore very
difficult to assign a {\em typical} Lorentz factor to the jet. 

%Models fitting the observed opening angle distribution typically fall
%into two distinct classes. The first matches the high initial opening
%angle close to the source and falls off steeply with increasing
%distance. At around ~100 \pc{} (unprojected distance) the over-expanded
%jet is recollimated towards the axis, causing the jet width to reach
%very small values before being deflected again from the axis. This
%causes a noticeable kink in the opening angle curve. The second class
%starts with lower initial opening angle and matches the observed
%opening angle smoothly along the length of the jet and does not show
%any recollimation shock (see \cite{GTB05}, Fig. 2).

In the following we discuss two exemplary models. We note
however, that due to the degeneracy of the parameter space, each model
could be realized by a range of formal parameter sets. Therefore, we
will not give specific parameter values and designate them simply
model 1 and model 2.

Figure \ref{fig:opening_angle} compares the opening angle as defined
by the separating fieldline $\Psi_\alpha$ with the observational
measurements. All our models fail to fit the jet width at large
distances close to the optical knot A at $\sim 900 \pc$ \cite{GTB05}.
Models with large initial opening angle, as model 1, have a hard time
reproducing the opening angle distribution. It is in general very
difficult to make the curve more concave. Models with initial opening
angles falling slightly short of the measured value, as eg model 2,
may easily reproduce the rest of the measurement and yield quite good
fits. Note, that both models show a clear kink at $\sim 70
\,\pc$. This is due to fact, that these jets, as most models,
over-expand and are forced to recollimate towards the axis.

\section{Synchrotron maps}

%Figures: 
%1) opening angle fit for model 1 and model 2
%2) synchrotron maps for model 1 and 2
%3) flux along axis for model 1 and 2 
%4) 3 profiles for model 1 and model 2, profiles separated by +1 or multiplot

\begin{figure}[t]
\center
\centerline{\psfig{figure=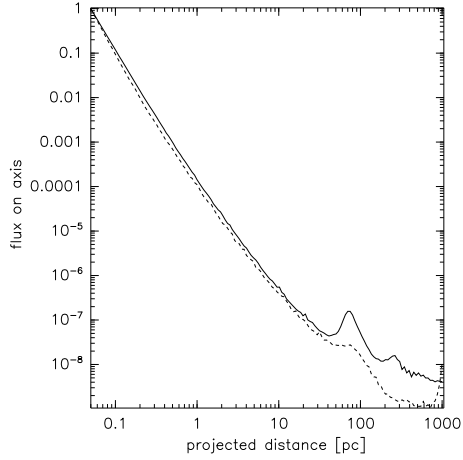,width=6truecm}}
\caption{Comparison of the synchrotron flux along the jet axis calculated from 
model 1 ({\em solid lines}) and model 2 ({\em dashed lines}),  respectively.}
 \label{fig:trend}
\end{figure}

\begin{figure}[t]
\center
\centerline{\psfig{figure=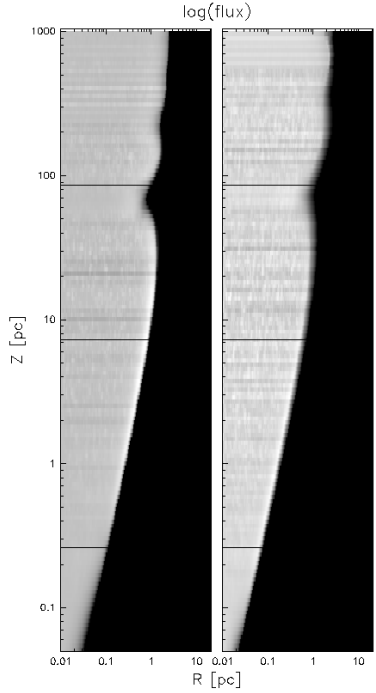,width=6truecm}}
\caption{Synthetic synchrotron map for model 1 ({\em left panel}) and
model 2 ({\em right panel}).}
\label{fig:map}
\end{figure}

\begin{figure}[t]
\center
\centerline{\psfig{figure=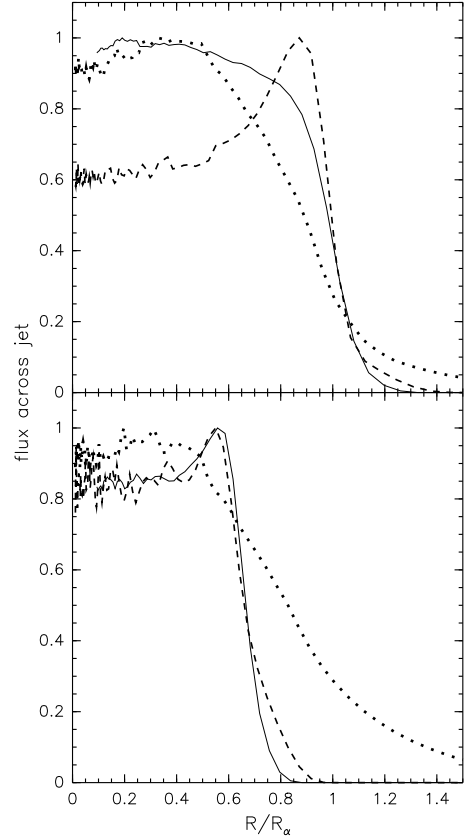,width=6truecm}}
\caption{Comparison of the synchrotron flux across the jet beam
calculated from model 1 ({\em upper panel}) and model 2 ({\em lower
panel}). Each panel shows profiles across the jet beam at three
different positions along the jet, ie close to the core ({\em thin
solid line}), at intermediate distance ({\em dashed line}), and at
large distance ({\em heavy dotted lines}). The lateral coordinate is
normalized to the width of the jet $R_\alpha$ as defined by the
dividing fieldline $\Psi_\alpha$.}
\label{fig:profile}
\end{figure}

The synthetic synchrotron emission maps shall qualitatively reproduce
1) opening angle defined through HWQM (half-width quarter-maximum), 2)
bright knot at the position of HST-1 at projected distance $\sim 70 \,\pc$ 3)
general morphology perpendicular to the jet axis, ie {\em limb-bright}
far from the core and {\em center-bright} close to the source, 4)
general trend along the axis $L_\nu \sim Z^{-1.5}$ \cite{Stawarzpriv}.

%Qualitatively reproduce: 
%\begin{itemize}
%    \item opening angle defined through HWQM
%    \item bright knot HST-1 at $\sim 100 {\rm pc}$\\
%      recollimation shock (Stawarz et al 2006)
%    \item {\em center-bright} profiles across the jet near the radio core
%    \item {\em limb-bright} profiles far away from the radio core
%    \item trend along jet axis, $L_\nu \sim Z^{-1.5}$
%  \end{itemize}
%  Be {\bf consistent} with observed $\sin i$

For the moment the calculation of the radio emissivity $\epsilon$ is
done following \cite{L81}. It is assumed,
that the radiating region is optically thin with a uniform and 
isotropic distribution of electrons
\begin{equation}
  N(E) \propto E^{-(2\alpha+1)}
\end{equation}
giving rise to radiation with spectral index $\alpha$, ie. $S_\nu
\propto \nu^{-\alpha}$ in the comoving frame.
%Note, that these quantities are measured in
%the comoving frame of the plasma. The maximum possible degree of
%polarization $p_0 = (3\alpha+3)/(3\alpha+5)$ is reached for a
%uniform magnetic field $\vec B$. 

%The fractional polarization $p$ can less than its maximum value $p_0$
%because of 1) small scale, gaussian magnetic field fluctuations which
%contribute to $B^2$ but cancel out, with respect to $\vec B$, when
%integrated over a small volume \cite[but see][]{LCB06}, and 2) a
%symmetry in the magnetic field along the line-of-sight, eg. pointing
%north in the foreground and south in the background.

%We assume further, that the fraction of electron number density to
%proton number density is constant throughout the emitting volume. Then
%the number density of electrons $n_e = \int N(E) \, dE$ is proportional to
%the density of the plasma. 
%, ie. $n_e \propto \rho$ in every frame of reference.

If the radiating plasma moves at relativistic speeds, ie
$\Gamma \gg 1$, the synchrotron emission has to
be evaluated in the comoving frame $\Sigma'$, instead of the lab- or
observers frame $\Sigma$. 
%All primed $'$ quantities refer to the 
%comoving frame. Note however, that the relative velocity $\vec v
%\itprod \los$, between the lab frame and the comoving frame is not constant
%within the emitting region, unless the velocity field $\vec v$ is
%homogeneous, which is not assumed. 
The synchrotron emissivity in the
comoving frame is given by the magnitude of the magnetic field
perpendicular to the line-of-sight and the
electron density as
%\begin{equation} \label{eq:em_c}
$  \epsilon' \propto \rho' |\vec B' \cross  \los'|^{\alpha+1}$,
%\end{equation}
where $\rho'$,$\vec B'$ and $\los'$ are the electron density,
magnetic field and line-of-sight unit vector in the comoving frame, respectively.
%are
%given in terms of the observers frame quantities through the
%Lorentz-transformations
%\begin{equation} \label{eq:denLT}
%\rho' =  \rho/\Gamma, 
%\end{equation}
%\begin{equation} \label{eq:BLT}
%\vec B' = \frac 1{\Gamma} \vec B 
%    + \frac{\Gamma}{\Gamma+1} \frac{\vec v}{c^2}(\vec v \itprod \vec B),
%    \qquad \vec E' = 0
%\end{equation}
%\begin{equation} \label{eq:nLT}
%  \los' = \dop \los - (\dop+1)\frac{\Gamma}{\Gamma+1}
%  \frac{\vec v}{c}. 
%\end{equation}
%We introduced the Doppler factor ${\cal D} = (\Gamma(1-\vec v \itprod
%\los/c))^{-1}$ for a compact notation and exploited the fact, that for
%ideal MHD Ohm's law holds as $\vec E = -\vec v \cross \vec B$

The emissivity  in the lab frame $\epsilon$ appears
Doppler-boosted as
%\begin{equation} \label{eq:em}
$  \epsilon = {\cal D}^{\alpha+2} \, \epsilon'$,  
%\end{equation}
with a Doppler factor ${\cal D} = (\Gamma(1-\vec v \itprod
\los/c))^{-1}$.
Finally, the flux in the plane of the sky, $I$,  is given by
integration along the line-of-sight.
%\begin{equation} \label{eq:I}
%  I = \int_{\rm los} \epsilon \, d\los.
%\end{equation}
%For small velocities, this relativistic expressions reduce to the
%non-relativistic as expected.
The amount of relativistic beaming strongly depends on the
line-of-sight angle, ie the angle between the jet axis and the
direction to the observer, which we fix to $\theta\ilos = 40\degr$.

In the following we discuss the same two exemplary models 
as in the previous section. 
%These will
%be measure against the objectives for the synchrotron maps presented
%earlier. 
To this end, we plotted spatial 2-dimensional synthetic
synchrotron emission maps (figure \ref{fig:map}), the flux measured
on the jet axis as a function of distance from the core (figure
\ref{fig:trend}), and the flux profiles across the jet at various
positions along the jet axis (figure \ref{fig:profile}).

Figure \ref{fig:trend} illustrates the trend along the axis.
% through the synchrotron flux measured on the axis as a function of projected
%distance.
Both models show similar power-law indices.  However, the
model jets dim out faster than the observed jet (power-law index $\sim
-2$ versus $-1.5$). Further, both models show a sharp rise of
luminosity around 70 \pc{}, where the flux increases by more than an
order of magnitude over the power-law. The location of this bright
spot coincides with the location of strong recollimation shocks
%, where density and magnetic field strength increase.

Figure \ref{fig:map} shows synthetic synchrotron maps. In order to
increase the contrast we divided the map by the trend along the jet
axis. The jet beam is well defined, showing large opening angles close
to the core and becoming almost conical a large distance.  Both models
show a strong decrease of the jet width forming a visible neck at the
position of the recollimation shock.

The synthetic synchrotron maps show strong gradients of flux across
the jet. These are more clearly visible in figure \ref{fig:profile},
were cuts across the jet are shown. 
%plotted at three different positions along the jet for both models. 
For model 1 the profiles clearly show limb-brightening from
distance $\sim 0.5 \, \pc$ to close to the neck in the jet. The
regions close to the core and the neck are brightest on the jet axis,
ie centrally-peaked. In the limb-bright region, the flux on axis
can be as small as half the limb flux. Model 2 generally shows less
variation of the flux across the jet. While close to the core the
emission is still centrally-peaked, limb-brightening takes over much
faster at $\sim 0.2 \, \pc$ but is not as prominent. The on-axis flux
is $\sim80\%$ of the limb.

\section{Summary}

%\cite{GTB05} used the shape of the fieldline separating the two region
%on the boundary surface to define the opening angle. This particular
%fieldline and flux contours of synthetic maps are parallel for a wide
%range of models and distances along the jet. However, the first method
%may over-estimate the opening angle for some parameters, as eg seen
%for model 2. 

The morphological structure across the jet can be reproduced in
principle. A wide range of models show limb-brightening away from the
core. However, this may be hidden due to projection and beaming effects for 
small line-of-sight angles.

Most models with acceptable fits to the opening angle distribution
feature a recollimation shock at the distance of HST-1. Only models
with very small initial opening angles do not seem to over-expand.

However, it is seem to be difficult to reconcile all observational
constraints with a single set of parameters. So far, no model of our
parameter study reproduces all features with high accuracy. This is
particularly true for models with a line-of-sight angle $\theta\ilos =
20\degr$. it seems to be easier at larger angles $\theta\ilos =
40\degr$. Clearly this calls for a denser and wider coverage of the
available parameter space.

Including of synchrotron total flux into the analysis in addition
to the openening angle distribution does not completely break the
degeneracy of the parameter space. Hopefully, extension to
polarization measurements will improve the situation.

\balance

\end{document}